\documentclass[sigchi]{acmart}

\AtBeginDocument{%
  \providecommand\BibTeX{{%
    \normalfont B\kern-0.5em{\scshape i\kern-0.25em b}\kern-0.8em\TeX}}}

\setcopyright{none}
\acmConference[CHI '19 Workshops]{  ACM SIGCHI International Conference on Human Factors in Computing Systems}{May 2019}{Glasgow, UK}
\acmISBN{978-1-4503-9999-9/18/06}

\begin{document}

\title{Championing Research Through Design in HRI}

\author{Michal Luria}
\email{mluria@cs.cmu.edu}
\affiliation{%
  \institution{Human-Computer Interaction Institute}
  \city{Carnegie Mellon University}
}

\author{John Zimmerman}
\email{johnz@cs.cmu.edu}
\affiliation{%
  \institution{Human-Computer Interaction Institute}
  \city{Carnegie Mellon University}
}

\author{Jodi Forlizzi}
\email{forlizzi.cmu.edu}
\affiliation{%
  \institution{Human-Computer Interaction Institute}
  \city{Carnegie Mellon University}
}

\renewcommand{\shortauthors}{Luria et al.}
\begin{abstract}
One of the challenges in conducting research on the intersection of the CHI and Human-Robot Interaction (HRI) communities is in addressing the gap of acceptable design research methods between the two. While HRI is focused on interaction with robots and includes design research in its scope, the community is not as accustomed to exploratory design methods as the CHI community. This workshop paper argues for bringing exploratory design, and specifically Research through Design (RtD) methods that have been established in CHI for the past decade to the foreground of HRI. RtD can enable design researchers in the field of HRI to conduct exploratory design work that asks \textit{what is the right thing to design} and share it within the community. 
\end{abstract}



\maketitle

\section{Introduction}
We agree with the workshop organizers that there are many challenges in working across the CHI and Human-Robot Interaction (HRI) communities. This workshop paper addresses the challenge of using exploratory design methods that are common in CHI to conduct design research on HRI topics. In particular, we discuss the method of Research through Design (RtD). 

RtD has been established in CHI after much effort from designers in the community. Their argument was that through the process of designing and making, designers can generate new knowledge and contribute to the research community~\cite{zimmerman2007research,koskinen2011design}. Today, RtD has become a valid form of inquiry and has a critical role in the design of human-computer interactions. One of the benefits of RtD is that it looks into ``making the right thing'', as opposed to ``making the thing right''~\cite{zimmerman2007research,buxton2010sketching}. In other words, it allows to research \textit{what} to design, and not what is the best way to design something that might have little value.

In the field of Human-Robot Interaction (HRI), much of the design research is focused on \textit{how to make a thing right}. With some notable exceptions, the majority of design papers published in HRI create and test a particular robot task or function. If the robot was successful in performing and communicating that particular task, then the design is successful. This approach is very similar to the approach that was common in the CHI community before designers established RtD methodology. 

We argue that the HRI community could greatly benefit from adopting exploratory design research methods from CHI. By doing so, the community could expand the boundaries of HRI and robot design, and gain knowledge about when designing robots is the right thing to do.

\begin{figure}
  \centering
  \includegraphics[width=0.75\columnwidth]{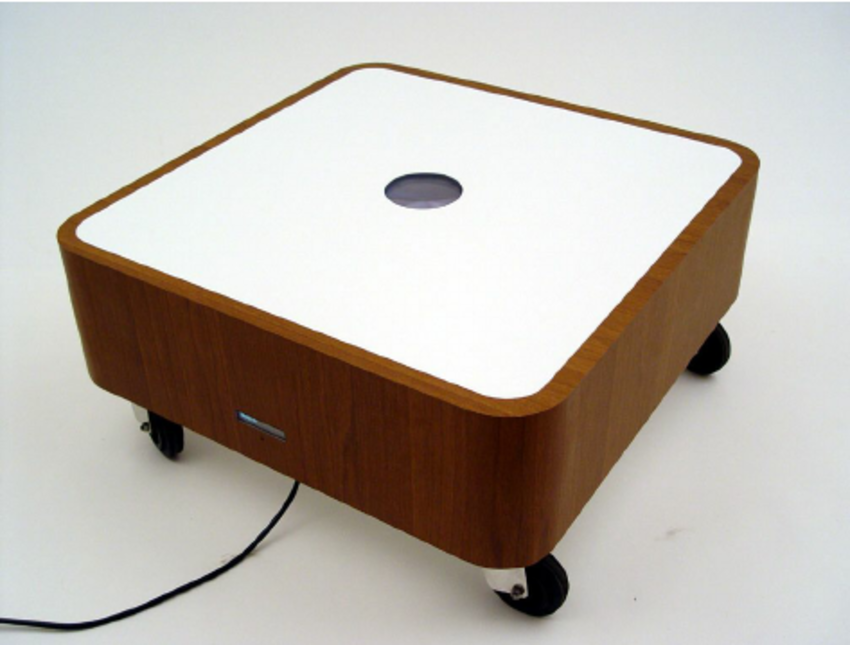}
  \caption{The Drift Table explores opportunities for ``ludic interaction'' with technology in the home. The table presents aerial maps that move according to the weight placed on the table. The prototype looks at curiosity and playfulness in technology for the home environment [7].}~\label{fig:drift}
\end{figure}

\section{Research Through Design}
One of the important aspects of Research through Design (RtD) and other methods that make use of design thinking is the ability to reconsider underlying assumptions. Previous work shows that in a community, new ideas are likely to adhere to elements from previous solutions and cause \textit{fixation} to solutions that are not useful in new contexts~\cite{chrysikou2005following}. 

For instance, early work in CHI presented \textit{Whisper}, a wearable device that allowed one to insert their fingertip into their ear canal to communicate in a phone call~\cite{fukumoto1999whisper}. Although the presented technology was very impressive and provided a technical contribution, the design was somewhat fixated on the assumption that people would not want to walk empty handed and seem as if they were ''taking to themselves''. In the perspective of time, we learn that this assumption was incorrect, and that people do not mind talking on the phone without a visible device. This realization brings a new range of designs and possibilities. Through exploratory design methods that re-frame and challenge assumptions, researchers can help avoid fixation and expand the boundary of acceptable designs in a field. 

An example of a RtD project from the CHI community is \textit{The Drift Table}~\cite{gaver2004drift}. The drift table is a coffee table that displays aerial photography in a small window at its center, according to the distribution of weight on it (see Fig.~\ref{fig:drift}). Through the process of designing the artifact, the authors explored how people perceive technology for the home and what are the opportunities to design for ``ludic experiences''~\cite{gaver2004drift}. Another example is Odom \textit{et al.}'s work on teenagers' virtual possessions. The authors conducted a series of exploratory interactions with teenagers, observed their relationships with physical and virtual possessions, and examined the design space of designing virtual possessions to be more meaningful~\cite{odom2011teenagers,odom2012fieldwork}. 

\begin{figure}
  \centering
  \includegraphics[width=1.0  
  \columnwidth]{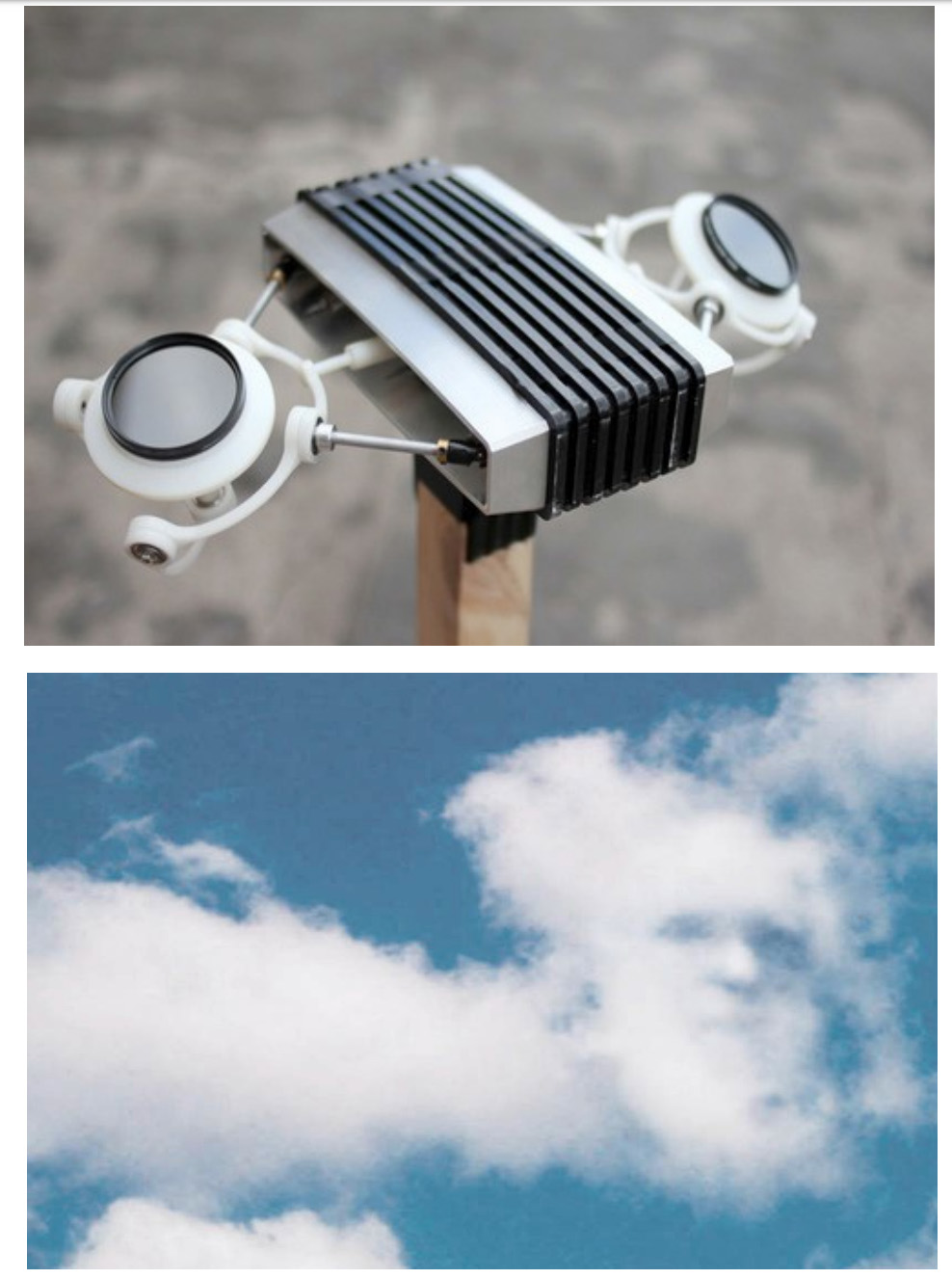}
  \caption{Auger used speculative design to explore why robots are not adopted as domestic objects, and suggested some alternatives. For example, a robot that looks at the clouds and notifies the user when there is a human-like face passing by [1].}
  \label{fig:speculativerobot}
\end{figure}

\begin{figure*}
  \centering
  \includegraphics[width=1.8\columnwidth]{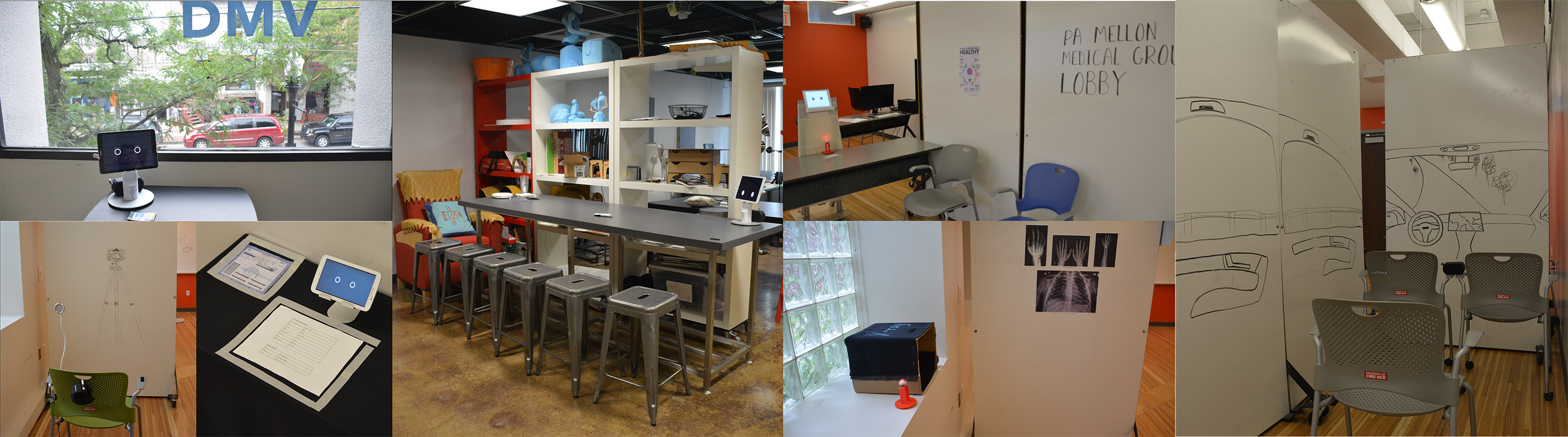}
  \caption{We used RtD to explore when robots' social presences might move from body to body, share a body, or each have a body of their own. Through Speed Dating, we tested a set of selected environments and contexts (a home, a car, a hospital and a public service).}
  \label{fig:reem-settings}
\end{figure*}

In the field of HRI, Auger used a Speculative Design approach~\cite{dunne2013speculative} to explore why robots are not becoming domestic products, and what might help them become such~\cite{auger2014living}. One robot presented in the paper observes clouds and recognizes human faces in them (see Fig.~\ref{fig:speculativerobot}). Through the interaction, the robot creates an emotional and playful connection with the user. Another project created \textit{IdleBot}, a robot that moves, but is barely interactive. This design questions whether and when robots should be engaging~\cite{overgoor2018idlebot}.

\section{The Challenge of Using RtD in HRI}
We recently conducted a RtD study that explored scenarios with multiple robots. We were interested in the question of when should robots' social presence (their presented entity and personality) move from one body to another (we call this action \textit{re-embodiment}), and when should the user be presented with a new personality altogether~\cite{luria2019reembodiment}. Our goal was to gain knowledge about the design space of sequential interaction with robots and agents across multiple locations and over time. While previous work looked at whether people perceive the movement of a social presence from one body to another~\cite{duffy2003agent}, it did not examine in what contexts and uses is this behavior valuable.

As this is a complex space of exploration that needs to probe interactions that do not yet exist, simply testing them in a lab setting will not do. Thus, we turned to the exploratory RtD method of Speed Dating with User Enactments~\cite{davidoff2007rapidly}. Using this method we emerged participants into scenarios that are ``brought-to-life'' using actors, prototypes and props to allow participants firsthand experiences of possible situations in the near future. Participants were then interviewed about their experiences. Just like romantic speed dating, participants got 'a sip' of many different scenarios through user enactments. By the end of the experience, they might not have learned much about any single scenario, but they are more likely to have a better sense of their own needs, desires and values on the topic. 

The main challenge of introducing such work in the HRI community is that the method is not structured or controlled. It is subjective and requires the researchers to frequently make design judgments, which is quite typical for RtD. Reviewing this work from a ``user study'' perspective is likely to point out plenty of methodology gaps, yet the goal of this type of work is not to systematically cover the entire space or reach internal validity. RtD is unstructured by definition---the flexibility allows exploration of an unknown design space and complicated ``wicked'' problems, in which the contexts and design choices are unlimited~\cite{buchanan1992wicked}. 

We currently share this work in the CHI and DIS communities, as they are more familiar with RtD approaches and this methodology. However, the \textit{topic} of research is more likely to be of interest to the HRI audience. In the near future, we would like to examine RtD for HRI to better understand how it would fit into the field, and how to encourage HRI researchers to accept RtD as a valid design contribution. We hope that doing so will enable researchers to conduct design work that questions underlying assumptions and expands the boundaries of HRI.

\section{Conclusion}
RtD allows designers and researchers to discuss \textit{what is the right thing to design}. This question is not frequently asked in the field of HRI, but might be a critical one for the development of its design perspective. As a field, we can use RtD to understand what is a robot, and what are the areas of exploration for designing them.

As the CHI community has already made significant efforts towards accepting RtD as a valid design contribution, instead of starting the argument from scratch, HRI designers can join forces with CHI designers to promote RtD in HRI. One of the ways to do this is to discuss what RtD methods might be more relevant for HRI, and what are the contributions they could bring. This can also help designers in HRI understand if there are adaptations that need to be made to RtD methods for HRI. Finally, publishing RtD on HRI topics in CHI is likely to create a growing group of researchers that are on the intersection of the fields, and allow RtD to enter HRI more naturally.

\balance{} 

\bibliographystyle{ACM-Reference-Format}
\bibliography{sample}

\end{document}